\documentclass[%
 reprint,
 amsmath,amssymb,
 aps, physrev,
]{revtex4-2}

\usepackage{graphicx}% Include figure files
\usepackage{dcolumn}% Align table columns on decimal point
\usepackage{bm}
\usepackage{amsmath}% bold math
\usepackage{multirow}
\usepackage{xcolor}
\usepackage{hyperref}% add hypertext capabilities
\begin{document}

\preprint{APS/123-QED}

\title{\textbf{Minimal model of sticky collisions in modulated optical traps} } 

\author{Hubert J. J\'{o}\'{z}wiak}
\email{hubert.jozwiak@ru.nl}
\affiliation{Institute for Molecules and Materials, Radboud University, Nijmegen, The Netherlands}
\author{Tijs Karman}
\email{t.karman@science.ru.nl}
\affiliation{Institute for Molecules and Materials, Radboud University, Nijmegen, The Netherlands}

\date{\today}% It is always \today, today,

\begin{abstract}
{Collisional loss of ultracold bialkali molecules is predicted to occur through photo-excitation of collision complexes.
However, recent measurements in modulated optical traps report an unexpected lack of excess survival,
at odds with theoretical predictions and earlier experiments.}
We develop a minimal quantum scattering model that treats collisions as a coherent process occurring in the presence of the modulated trap.
We demonstrate that quantum reflection from the long-range potential fundamentally alters loss dynamics, removing the tension between recent experiments and predicted sticking times.
{This framework provides new testable predictions such as a strong dependence on collision energy, initial molecular state, and long-range interactions, that are absent in the currently used phenomenological model.}
\end{abstract}

\maketitle

%\tableofcontents

Ultracold, ground-state bialkali molecules suffer from rapid depletion from optical traps via two-body collisions~\cite{Ospelkaus_2010, Takekoshi_2014, Guo_2016, ye2018collisions, Guo2018, Gregory_2019, Voges_2020}. 
Strikingly, this occurs even for non-reactive species~\cite{Zuchowski2010}, where the observed loss rates frequently approach the universal limit~\cite{Idziaszek2010}. 
This limit implies that essentially every collision reaching the short-range interaction region leads to the loss of both molecules. 
While this can now be circumvented by engineering repulsive long-range shields with external fields~\cite{Avdeenkov_2006, Buchler2007, Gorshkov_2008, cooper:2009, Karman2018, Lassabliere2018, Karman2025}, the underlying physical mechanism driving this two-body loss remains {the} unresolved question in ultracold chemistry.

The leading hypothesis attributes {loss} to the formation of transient collision complexes in so-called \textit{sticky collisions}~\cite{mayle2012statistical, Mayle2013,bause2023ultracold}. 
Although the molecules collide at ultracold temperatures ($< 1\,\mu\text{K}$), the interaction strength at short range is at least nine orders of magnitude larger. 
This deep potential well provides access to an immense density of states $\rho$, resulting in chaotic motion of the intermediate complex. 
Assuming fully ergodic dynamics, the complex lifetime can be evaluated using the statistical Rice-Ramsperger-Kassel-Marcus (RRKM) theory, yielding  $\tau_{c}^{\mathrm{RRKM}} = 2\pi\hbar\rho / N_{\mathrm{open}}$. 
Here, $N_{\mathrm{open}}$ is the number of asymptotically open channels, {which} equals $1$ for non-reactive species in their absolute ground state. 
{This gives rise to microsecond sticking times that are orders of magnitude longer than the picosecond timescales typical for simple reaction intermediates~\cite{Christianen2019PRA}.}
{During these prolonged lifetimes, complexes can be electronically excited by the optical dipole trap (ODT) laser, causing irreversible loss~\cite{Christianen2019a}.}

\begin{figure}
    \centering
    \includegraphics[width=\linewidth]{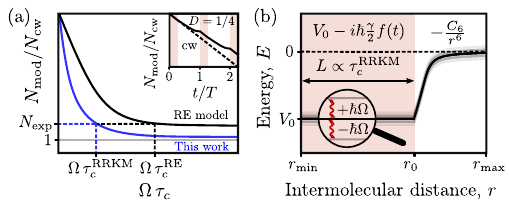}
    \caption{(a) {Inset:} Principle of chopped ODT experiments. The laser is on during the shaded fraction $D$ of the cycle. {Dark periods} $(1-D)$ allow transient complexes to dissociate, yielding a macroscopic excess survival of molecules (solid line) compared to a continuous wave (cw) trap (dashed). 
    {Main panel: Excess survival $N_{\mathrm{mod}}/N_{\mathrm{cw}}$ evaluated within the RE model (solid black line), demonstrating its dependence on the scaling parameter $\Omega\tau_{c}$.}
    {(b) Proposed quantum scattering model. The interaction combines a long-range vdW tail ($-C_6/r^6$) with a flat short range tuned to yield $\tau_{c} = \tau_{c}^{\mathrm{RRKM}}$.}
    Photoexcitation of the complex is modeled as a  {time-periodic} absorbing potential $-i\hbar\frac{\gamma}{2}f(t)$ coupling Floquet sidebands (faded lines) strictly at short range.}
    \label{fig:cartoon}
\end{figure}

Transient complexes have been directly detected in reactive KRb collisions~\cite{Hu2019}, {with complex lifetimes and photoexcitation rates consistent with theoretical predictions~\cite{Liu_2020}}. 
In parallel, the photoexcitation hypothesis in non-reactive bialkalis has been investigated using time-modulated (chopped) ODTs~\cite{Gregory2020, Gregory_2021, Gersema2021} and blue-detuned dark optical traps~\cite{Bause2021}. 
In chopped ODTs, the trapping laser is modulated with a duty cycle $D$ at a frequency $\Omega$ [inset in Fig.~\ref{fig:cartoon}(a)], while maintaining the same time-averaged intensity as a continuous-wave (cw) trap. 
This introduces dark periods that allow transient complexes to form and dissociate without suffering photo-induced loss, which should lead to a measurable excess survival. 
While initial chopped ODT experiments on $^{87}$Rb$^{133}$Cs observed loss suppression in reasonable agreement with the RRKM theory~\cite{Gregory2020}, subsequent measurements in $^{23}$Na$^{87}$Rb, bosonic $^{23}$Na$^{39}$K~\cite{Gersema2021}, fermionic $^{23}$Na$^{40}$K~\cite{Bause2021} found no such suppression. 
Furthermore, additional measurements on RbCs revealed an unexpected dependence on the initial hyperfine state~\cite{Gregory_2021}. 
{To reconcile the lack of observed loss suppression with the photoexcitation hypothesis, the true complex lifetime was concluded to exceed RRKM predictions by orders of magnitude~\cite{Gersema2021,Bause2021,bause2023ultracold}}.

Crucially, however, this conclusion depends entirely on the theoretical framework used to extract the sticking time $\tau_c$ from the observed excess survival. 
Despite extensive theoretical work on various aspects of sticky collisions~\cite{Croft2020, Quemener2021, Man2022, Croft_2023, Xu2026, croft2026longrangestatescollisionsultracold},
no quantum scattering theory directly predict the actual observable in these experiments: loss rates in time-modulated optical traps.
Instead, the experiments deduce $\tau_{c}$ via a phenomenological rate equation (RE) model.
This approach treats molecules and complexes as effectively separate species, with their respective densities ($n_{m}$ and $n_{c}$) governed by
\begin{equation}
\label{eq:re_model}
\dot{n}_{m} = -k^{\mathrm{cw}}_{\mathrm{loss}} n_m^2 + 2\frac{n_c}{\tau_{c}}, \quad \dot{n}_c = \frac{1}{2} k^{\mathrm{cw}}_{\mathrm{loss}} n_m^2 - \frac{n_c}{\tau_{c}} - \gamma(t) n_c,
\end{equation}
where $k^{\mathrm{cw}}_{\mathrm{loss}}$ is the complex formation rate (taken as loss rate measured in cw trap) and $\gamma(t)$ is the laser-induced destruction rate.
Assuming the laser intensity is sufficient to instantly destroy any complex ($\gamma \gg \tau_c^{-1}$), the effective loss rate $k^{\mathrm{RE}}_{\mathrm{loss}}$ in the modulated ODT is
\begin{equation}
\label{eq:re_model_rate}
    k^{\mathrm{RE}}_{\mathrm{loss}} = k_{\mathrm{loss}} \left[ D + \frac{\Omega \tau_{c}}{2\pi}  \left( 1 - e^{-\frac{2\pi (1-D)}{\Omega \tau_{c}}} \right) \right].
\end{equation}
The observable excess survival, $N_{\mathrm{mod}}/N_{\mathrm{cw}} = (1 + \bar{n}_0 k_{\mathrm{loss}} t)/(1 + \bar{n}_0 k^{\mathrm{RE}}_{\mathrm{loss}} t)$ for a hold time $t$, is determined by the dimensionless scaling parameter $\Omega \tau_{c}$.

In this Letter, we introduce a minimal model for sticky collisions in a modulated optical trap.
We treat the chopped ODT as a time-periodic complex absorbing potential at short range [Fig.~\ref{fig:cartoon}(b)]. 
Unlike the RE model, {this framework} describes collisions in the modulated trap as a coherent process,
including interference {effects sensitive} to {the} short-range phase shift and quantum reflection off the long-range interaction potential.
Such coherent dynamics introduce a simultaneous dependence of the observed loss on the temperature, 
the characteristic van der Waals (vdW) energy $E_{\mathrm{vdW}}$,
and modulation frequency, which is absent in the RE model.
This affects the dependence of the excess survival on the modulation frequency,
which can lead to a drastically different inferred sticking time $\tau_c$, see Fig.~\ref{fig:cartoon}(a).
{We demonstrate that our approach can reproduce the experimental observables without requiring the sticking time to be orders of magnitude larger than RRKM predictions, resolving the apparent tension between experiment and statistical theory.}

We model collisions of structureless molecules scattering along a relative distance coordinate $r$. The dynamics are governed by the time-dependent Hamiltonian
\begin{equation}
\hat{H}(t) = -\frac{\hbar^2}{2\mu}\frac{\partial^2}{\partial r^2} + \frac{\hbar^{2}l(l+1)}{2\mu r^{2}} + V(r) - i\hbar\frac{\gamma(r,t)}{2},
\end{equation}
where $\mu$ is the reduced mass of the system and $l$ is the partial wave quantum number. The interaction potential has the vdW form at long range ($r > r_{0}$), $-C_6/r^6$, a constant short range,  {$V_{0}=-C_{6}/r_{0}^{6}$,} 
and a hard wall at $r_{\mathrm{min}}$, see Fig.~\ref{fig:cartoon}(b).
We choose $r_\mathrm{min} \approx r_0-\tau_c^\mathrm{RRKM}\sqrt{|V_0|/2\mu}$,
such that the classical round-trip time matches the RRKM sticking time, $\tau_{c}^{\mathrm{RRKM}}$ \footnote{Since, for bialkalis, $\tau_{c}^{\mathrm{RRKM}} \sim 10^{-4}\,\text{s}$ (Table~\ref{tab:molecules}), matching the sticking time requires placing hard-wall wall at extremely large negative values, $r_{\mathrm{min}} \sim -10^8\,a_0$.},
while small variations of $r_\mathrm{min}$ determine the short-range phase shift $\Phi = (r_0-r_\mathrm{min}) \sqrt{2\mu |V_0|} / \hbar$.
{The presence of $\Phi$ enables our model to capture interference between incoming and reflected waves, which is absent  the incoherent two-step RE model.}

Photoexcitation of the complex is modeled by a time-periodic absorbing potential, $-i\hbar \gamma(r,t)/2$,
where $\gamma(r,t) = \gamma_0 f(t)$ for $r \le r_0$ and zero elsewhere. 
The modulation profile $f(t)$ is a periodic square wave of duty cycle $D$ and period $T=2\pi/\Omega$, centered at $t=0$ 
(i.e. $f(t) = 1$ for $t \in [-DT/2, DT/2]$, and $0$ otherwise). 
The resulting observables are insensitive to the exact value of $\gamma_0$, provided it is large enough to operate in the saturated loss regime, yet sufficiently small ($\hbar\gamma_0 \ll |V_0|$) to avoid spurious reflection at $r_{0}$.
\color{black}

\begin{table}
    \centering
    \caption{Molecules considered in calculations. $C_{6}$ coefficients taken from~\cite{Zuchowski2013}. $\tau^{\mathrm{RE}}_c$ denotes sticking time determined experimentally using the RE model [Eq.~\eqref{eq:re_model}].}
    \begin{tabular}{lrrr} \hline \hline
         Molecule & $C_{6}$ (a.u.) & $\tau_{c}^{\mathrm{RRKM}}$  & $\tau_{c}^{\mathrm{RE}}$   \\ \hline
         $^{87}$Rb$^{133}$Cs & $180\,982$ & $0.253\,\text{ms}$~\cite{Christianen2019PRA}   & $0.53(6)\,\text{ms}$~\cite{Gregory2020} \\
         $^{23}$Na$^{87}$Rb &$1\,507\,089$ & $19\,\mu\text{s}$~\cite{Liu_2022} & $>1.2\,\text{ms}$~\cite{Gersema2021}\\
         $^{23}$Na$^{40}$K & \multirow{2}{*}{$516\,606$} & $18\,\,\mu\text{s}$~\cite{Liu_2020} & $> 1.4\,\text{ms}$~\cite{Bause2021}\\
         $^{23}$Na$^{39}$K &  & $6\,\mu\text{s}$~\cite{Liu_2020} & $> 0.35 \,\text{ms}$~\cite{Gersema2021}\\
         \hline \hline
    \end{tabular}
    \label{tab:molecules}
\end{table}

Because the Hamiltonian is periodic in time, we employ the Floquet formalism and expand the total scattering wavefunction into Fourier components
\begin{equation}
\label{eq:wv_exp}
\Psi(r,t) = e^{-iEt/\hbar} \sum_{n=-\infty}^{\infty} \psi_n(r) e^{-in\Omega t},
\end{equation}
where $E$ is the collision energy. 
We choose the incoming channel in the $n=0$ sector, with the asymptotic wavevector $k_0 = \sqrt{2\mu E}/\hbar$. Inserting Eq.~\eqref{eq:wv_exp} into the time-dependent Schr\"{o}dinger equation yields a set of time-independent coupled channel (CC) equations
\begin{equation}
\label{eq:cc}
   \frac{d^2}{dr^2}\mathbf{\Psi}(r)  + \frac{2\mu}{\hbar^2} \Bigl(E - \mathbf{W}(r) \Bigr) \mathbf{\Psi}(r) = 0 .
\end{equation}
The $\mathbf{W}$ matrix is distance-independent at short range ($r\leq r_{0}$) and diagonal at long range ($r>r_{0}$).
At short range, the imaginary term couples the Fourier components via $W_{nm} = (V_{0}-n\hbar\Omega) \delta_{nm} - \frac{i\hbar}{2} \gamma_{n-m}$, where $\gamma_{k} = \gamma_0 D \, \mathrm{sinc}(k\pi D)$ is the Fourier transform of the modulation profile. 
At long range, the problem decouples into independent single-channel potentials with $W_{nn}(r) = -C_{6}/r^{6} + \hbar^{2}l(l+1)/(2\mu r^{2})$, which admit exact analytical solutions~\cite{Gao1998}.
The full quantum scattering problem is thus reduced to a matrix diagonalization at short range and single-channel scattering at long range, requiring only a single matching step at $r_0$ (see End Matter).
The energy-dependent loss rate coefficient is then extracted from the scattering matrix \footnote{{Neglecting interference terms between different Fourier components mathematically corresponds to evaluating a time-averaged loss rate over the modulation cycle.}} (see End Matter)
\begin{equation}
\label{eq:rate_our_model}
k_{\mathrm{loss}}(E) = \frac{2\pi \hbar}{\mu k_{0}} \left( 1 - |S_{0,0}(E)|^2 \right).
\end{equation}
\color{black}
To directly compare with experiments, we compute the thermally averaged loss rate $\langle k_{\mathrm{loss}} \rangle$ by integrating $k_{\mathrm{loss}}(E)$ over the Maxwell-Boltzmann energy distribution at temperature $T$. 
All numerical parameters of the model are rigorously tested to ensure convergence of the $k_{\mathrm{loss}}(E)$ to a few percent level (see End Matter).

With this framework, we perform loss rate calculations for the non-reactive bialkalis investigated in recent chopped ODT experiments~\cite{Gregory2020, Gregory_2021, Gersema2021, Bause2021}. The relevant molecular parameters are summarized in Tab.~\ref{tab:molecules}.

\begin{figure}
    \centering
    \includegraphics[width=\linewidth]{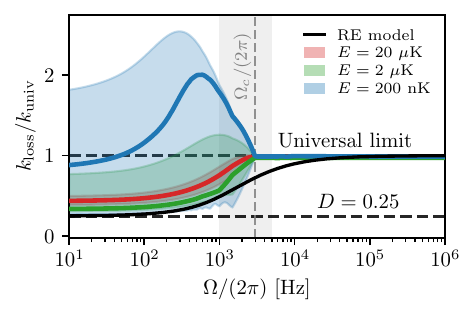}
    \caption{Breakdown of the RE model in ultracold collisions. 
    The computed $s$-wave two-body loss rate coefficient for $^{87}$Rb$^{133}$Cs ($D=0.25$), normalized to the universal limit $k_{\text{univ}} = \frac{8\pi\hbar}{\mu}\bar{a}$~\cite{Idziaszek2010}, is shown as a function of the modulation frequency $\Omega$. 
    The solid black line denotes the phenomenological RE prediction [Eq.~\eqref{eq:re_model_rate}] assuming $\tau_c = \tau_c^{\mathrm{RRKM}}$. 
    Colored shades indicate the full range of achievable loss rates obtained by varying the short-range phase $\Phi$, while solid lines show results for specific phases. 
    The gray region highlights the frequency window probed in experiments~\cite{Gregory2020, Gregory_2021}. 
    Energies correspond to the experimental temperature ($2\,\mu\mathrm{K} \approx E_{\mathrm{vdW}}/k_B$), and values an order of magnitude lower and higher. }
    
    \label{fig:loss_vs_chop}
\end{figure}

{Figure~\ref{fig:loss_vs_chop} compares the phenomenological RE model [Eq.~\eqref{eq:re_model_rate}] against our approach [Eq.~\eqref{eq:rate_our_model}] for $^{87}$Rb$^{133}$Cs.}
Within the RE model, the normalized loss $k_{\mathrm{loss}}^{\mathrm{RE}}/k_{\mathrm{univ}}$ is temperature-independent. 
It transitions from $D$ at slow modulation ($\Omega \tau_{c} \ll 1$) to $1$ (universal loss) at high frequencies ($\Omega \tau_{c} \gg 1$).
Experiments operating in the few-kHz window (gray band) expect observable loss suppression~\cite{Gregory2020,Gregory_2021,Gersema2021, Bause2021}.

For high collision energies ($E \gg E_{\mathrm{vdW}}$, red curve), our model qualitatively recovers the RE prediction, albeit with two fundamental differences. 
First, at low frequencies, our model yields $k_{\mathrm{loss}}/k_{\mathrm{univ}} > D$.
This is a consequence of the definition of survival, [Eq.~\eqref{eq:rate_our_model}],
which excludes transitions to other Floquet channels, thereby accounting for two-body heating that is absent in the RE model.
Whether two-body heating should be counted as loss depends on the experimental settings such as trap depth,
and alternative definitions are possible; see End Matter.
Second, at high frequency, our model reaches the universal loss rate limit sooner.
In the RE model, the complex decay is treated as a Poisson process, meaning a long-lived tail of complexes requires faster modulation to be completely absorbed.
In our single-channel model, molecules are exposed to the absorbing potential over a fixed round-trip time $\tau_c^{\mathrm{RRKM}}$,
such that $k_{\mathrm{loss}}$ saturates at a critical frequency $\Omega_{c}/2\pi = (1-D)/\tau_c^{\mathrm{RRKM}}$.
We note that neither distribution of $\tau_c$ is necessarily correct for ultracold collision complexes~\cite{Croft_2023}.

As the collision energy approaches $E \lesssim E_{\mathrm{vdW}}$, the normalized rate $k_{\mathrm{loss}}/k_{\mathrm{univ}}$ begins to depend on energy and the unknown short-range phase $\Phi$. 
The observed loss can then be smaller or larger than the universal rate observed in the cw trap, as indicated by the shaded regions in Fig.~\ref{fig:loss_vs_chop}. 
This sensitivity arises from quantum reflection at the long-range vdW tail: a molecule leaving the short range can reflect and access the absorbing potential multiple times. 
The resulting quantum interference between incoming and reflected waves can either enhance or suppress the loss. 
For particular values of $\Phi$, quantum reflection completely masks the expected loss suppression precisely in the experimental frequency window, as shown by the solid lines in Fig.~\ref{fig:loss_vs_chop}.
\color{black}

\begin{figure*}[!ht]
    \centering
    \includegraphics[width=\linewidth]{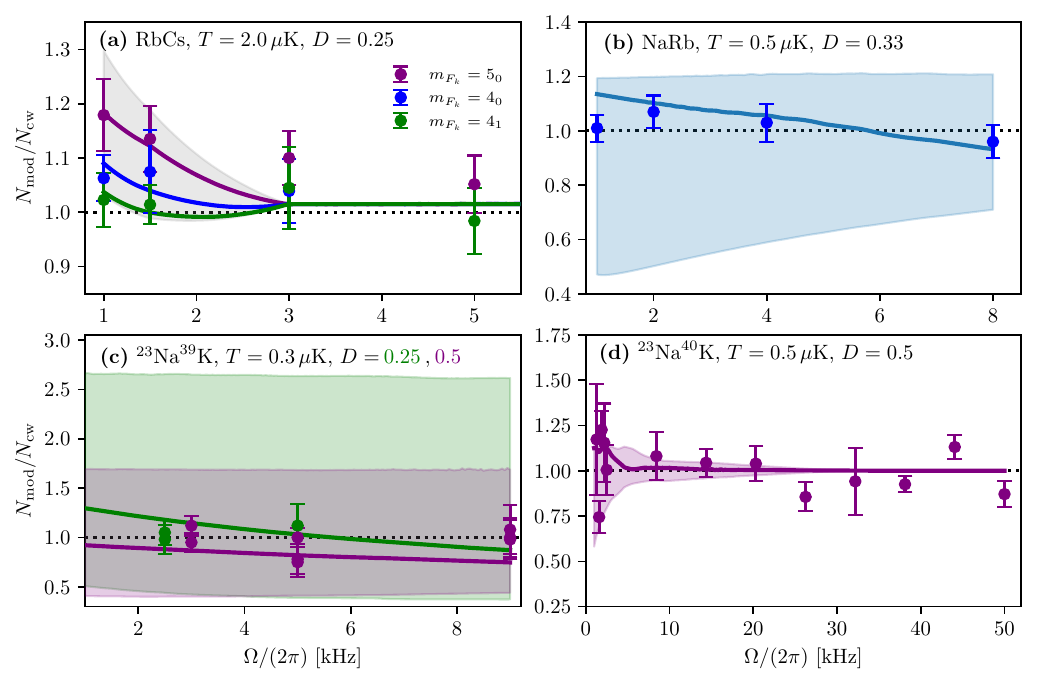}
    \caption{Excess survival $N_{\mathrm{mod}}/N_{\mathrm{cw}}$ in chopped ODTs across different non-reactive bialkali species. Experimental data (symbols) are compared against our model, which exhibits phase-dependent dispersion (shaded bands). We present specific short-range phases (solid lines) that match the experimental results. In all calculations, complex lifetime $\tau_{c}$, simulated as the classical time round-trip time spent at short-range, is fixed to the RRKM-predicted value (Table~\ref{tab:molecules}). \textbf{(a)} State-dependent survival of RbCs for three hyperfine states ($m_{F_{k}}$) in the rovibrational ground state manifold~\cite{Gregory_2021}. \textbf{(b)} $^{23}$Na$^{87}$Rb~\cite{Gersema2021}. \textbf{(c)} Bosonic $^{23}$Na$^{39}$K~\cite{Gersema2021}, evaluated at different duty cycles $D$. \textbf{(d)} Fermionic $^{23}$Na$^{40}$K~\cite{Bause2021}.}
    \label{fig3}
\end{figure*}

The short-range phase $\Phi$ is strictly defined by the specific quantum state of the colliding molecules. 
However, computing its exact value \textit{ab initio} is unfeasible, as this would require including all rovibrational and spin degrees of freedom~\cite{Croft_2017, Kendrick_2021, Morita2024, Liu2025, da_Silva_2025, Kendrick_2026}.
We therefore treat $\Phi$ as a parameter of the model, leveraging the sensitivity of $k_{\mathrm{loss}}$ to reinterpret recent chopped-ODT experiments~\cite{Gregory_2021, Gersema2021, Bause2021}.
\color{black}

To confront our model with experimental data, we compute the excess survival ratio {using the formula introduced below Eq.~\eqref{eq:re_model_rate}, substituting the phenomenological rates with computed thermally averaged values: $N_{\mathrm{mod}}/N_{\mathrm{CW}} = (1 + \bar{n}_0 \langle k_{\mathrm{univ}} \rangle t)/(1 + \bar{n}_0 \langle k_{\mathrm{loss}} \rangle t)$.}

Figure~\ref{fig3}(a) compares our model with {the excess survival measured for ultracold} $^{87}$Rb$^{133}$Cs molecules~\cite{Gregory_2021}.
In these experiments, molecules were prepared in three distinct hyperfine levels of the absolute rovibrational ground state.
Loss suppression was found to be state-dependent, being most pronounced for the lowest-energy state ({labeled} $5_0$).
{Using the phenomenological RE model, this observation was interpreted as a state-dependent sticking time, yielding $\tau_{c} = 8(3)\times 10^{-4}\,\mathrm{s}$ for the $5_0$ state}, roughly a factor of three longer than the RRKM prediction (Table~\ref{tab:molecules}).
The weaker suppression observed for excited hyperfine states was attributed either to even longer $\tau_{c}$ or to inelastic hyperfine-changing collisions in the dark.

Our model offers an alternative explanation. Assuming nuclear spins remain spectators during the collision ($N_{\mathrm{open}}=1$), distinct initial hyperfine states effectively sample slightly different short-range interaction potentials, manifesting as distinct values of $\Phi$. Because the computed dispersion of loss rates covers all experimental data, we can identify specific phase shifts [solid lines in Fig.~\ref{fig3}(a)] that reproduce the survival of each hyperfine state, all while enforcing the complex lifetime to be exactly equal to the RRKM-predicted value, $\tau_{c}^{\mathrm{RRKM}} = 253\,\mu\text{s}$.

We now apply this analysis to NaRb and bosonic NaK~\cite{Gersema2021}. 
Due to their lower density of states, these species have shorter RRKM lifetimes (Table~\ref{tab:molecules}). 
{Consequently, the $1-10\,\mathrm{kHz}$ modulation frequencies used in the experiments are within the slow-chopping regime ($\Omega \tau_c \ll 1$), where the RE model anticipates measurable loss suppression.} 
The observed lack of suppression was therefore interpreted as evidence for anomalously long complex lifetimes, exceeding RRKM predictions by up to two orders of magnitude~\cite{Gersema2021}.
However, because the experimental temperatures ($300\,\mathrm{nK}$ for NaK and $500\,\mathrm{nK}$ for NaRb) are far below their respective vdW energies ($E_{\mathrm{vdW}} \approx 11.6\,\mu\text{K}$ and $2.8\,\mu\text{K}$), reflection off the long-range tail significantly alters the scattering dynamics. 
As shown in Figs.~\ref{fig3}(b,c), the {resulting phase-dependent quantum interference yields a broad range of possible excess-survival curves} that fully covers the experimental data. 
We can identify specific phases that quantitatively reproduce the observations, including measurements taken at multiple duty cycles for NaK, all while using the RRKM-predicted complex lifetimes.

Next, we apply our model to fermionic NaK~\cite{Bause2021}, which exhibited no loss suppression {within the investigated frequency range of $1-50\,\text{kHz}$}.
For identical fermions, collisions are restricted to the $p$-wave ($l=1$) channel. 
{The RRKM sticking time is three times longer than for bosonic NaK (Tab.~\ref{tab:num_params}), shifting the critical modulation frequency for loss suppression to $\Omega_{c}/(2\pi) \approx 27\,\mathrm{kHz}$.
Above this value, no loss suppression is expected within our model, whereas at lower frequencies, our calculations reveal a phase-dependent dispersion of excess survival.}
We find short-range phases that reproduce the lack of suppression {across the entire frequency range,} again {while maintaining $\tau_{c} = \tau_{c}^{\mathrm{RRKM}}$}.

    \begin{figure}[!ht]
        \centering
        \includegraphics[width=\linewidth]{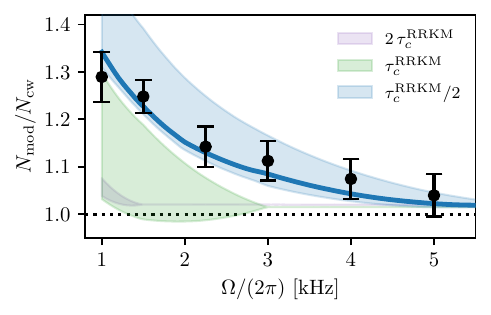}
        \caption{$\tau_{c}$-dependent excess survival in RbCs compared with the first chopped ODT measurement~\cite{Gregory2020}. Shaded bands show the phase-dependent dispersion of $N_{\mathrm{mod}}/N_{\mathrm{cw}}$ computed with our model under different assumptions of $\tau_{c}$. Solid line represents a specific short-range phase that reproduces experimental data best.}
        \label{fig4}
    \end{figure}

Finally, we revisit the first chopped ODT measurement on RbCs~\cite{Gregory2020}. 
{Our model explains why this experiment alone observed loss suppression: unlike others~\cite{Gersema2021, Bause2021}, it operated at $E \approx E_{\mathrm{vdW}}$, where quantum interference effects are less pronounced.}
Using the RE model, the authors originally extracted a complex lifetime roughly twice as long as the RRKM prediction (Table~\ref{tab:molecules}). 
As shown in Fig.~\ref{fig4}, evaluating our model with {$\tau_{c} = \tau_{c}^{\mathrm{RRKM}}$ yields no value of the short-range phase $\Phi$ capable of reaching the measured excess survival.} 
Furthermore, if we adopt the RE conclusion and artificially enforce $\tau_{c} = 2\,\tau_{c}^{\mathrm{RRKM}}$, the onset of the phase-dependent dispersion shifts to proportionally lower frequencies, making the disagreement even worse. 
Instead, we find that the experimental data can be explained within our model only by assuming a \textit{shorter} sticking time of $0.5\,\tau_{c}^{\mathrm{RRKM}}$.

We limit the application of our model to molecule-molecule collisions, discussed above,
but note that similar discrepancies have been observed in atom-molecule mixtures~\cite{Gregory_2021,Nichols_2022,bause2023ultracold}.
In these systems, the unpaired electronic spin may be strongly coupled and involved in the dynamics,
which is not accounted for in our single-channel model.

In conclusion, we have developed a minimal quantum scattering model for ultracold molecular collisions in time-modulated optical traps.
In contrast to phenomenological rate equations, which treat molecules and complexes as essentially separate species, our approach describes the entire collision as a single coherent event. 
By replacing the incalculable short-range complexity with a single phase-sensitive boundary condition, our model naturally incorporates wavefunction interference and quantum reflection off the long-range potential.
This establishes a dependence of the excess survival on the interplay between the collision energy $E$, the van der Waals scale $E_{\mathrm{vdW}}$, the modulation frequency $\Omega$, and the complex lifetime $\tau_c$.
Such a connection is inherently absent in RE models, which decouple the complex dynamics ($\tau_c, \Omega$) from the collision kinematics ($E_{\mathrm{vdW}}, E$).

Applying this framework to recent chopped-ODT experiments~\cite{Gregory2020,Gersema2021,Gregory_2021,Bause2021}, we show that the unexpected absence of loss suppression does not necessarily imply anomalously long complex lifetimes.
Rather, it can emerge from quantum reflection off the long-range barrier in the ultracold regime ($E \lesssim E_{\mathrm{vdW}}$). 
By incorporating these coherent effects, our model relieves the tension between experimental observations and statistical RRKM theory. 
Ultimately, we emphasize that extracting sticking times from modulated traps is inherently model-dependent.

Looking forward, our framework provides testable predictions that fall strictly outside the scope of the RE model.
First, tuning the collision energy $E$, or temperature $k_B T$, across the $E_{\mathrm{vdW}}$ threshold will alter the quantum reflection probability, modifying the response to trap modulation.
At fixed temperature, our model similarly predicts a dependence on the long-range interaction, which can be tuned for ultracold molecules.
This can be achieved without applying external fields that could affect sticking dynamics~\cite{Man2022} by preparing molecules in mixtures of $j=0$ and $j=1$ rotational states, which experience longer-ranged resonant dipolar interactions characterized by lower energy scales.
Furthermore, our results demonstrate that parameterizing sticky collisions exclusively by the molecular species is physically insufficient. 
Because the short-range phase $\Phi$ is determined by the specific quantum state and external fields, experiments probing the same molecule can yield drastically different loss rates.
While exact \textit{ab initio} predictions of $\Phi$ remain intractable, our model establishes theoretical bounds that can be rigorously tested.

\textit{Acknowledgments---}
The research was funded by the European Union (Project No.~101269084, HORIZON-MSCA-2025-PF, 2STICKY). The views and opinions expressed are, however, those of the authors only and do not necessarily reflect those of the European Union or the European Research Executive Agency. Neither the European Union nor the granting authority can be held responsible for them.
T.K.~is supported by NWO VIDI (grant ID 10.61686/AKJWK33335).

\section*{End Matter}
\textit{Appendix A: Details of numerical calculations.---} Since $\mathbf{W}$ in Eq.~\eqref{eq:cc} is distance-independent at short range ($r_{\mathrm{min}}\leq r \leq r_{0}$), we diagonalize it as $\mathbf{W} = \mathbf{U} \mathbf{\Lambda} \mathbf{U}^{-1}$, where $\Lambda_{mn} = \lambda_{m} \delta_{mn}$. 
With the hard-wall boundary conditions at $r_{\mathrm{min}}$, the short-range matrix solution in the adiabatic basis is
\begin{equation}
[\mathbf{\Psi}^{\mathrm{ad}}(r)]_{mn} = \sin(q_m(r - r_{\mathrm{min}})) \delta_{mn},
\end{equation}
where $q_m = \sqrt{2\mu(E - \lambda_m)}/\hbar$. 
We construct the ratio matrix at $r_{0}$ as $\mathbf{Q}^{\mathrm{ad}}(r_0) = \mathbf{\Psi}^{\mathrm{ad}}(r_0 - h) [\mathbf{\Psi}^{\mathrm{ad}}(r_0)]^{-1}$, where $h$ is the grid step size.
This matrix is then transformed to the asymptotic basis: $\mathbf{Q}(r_0) = \mathbf{U} \mathbf{Q}^{\mathrm{ad}}(r_0) \mathbf{U}^{-1}$.

At long range ($r > r_0$), the channels decouple, and the problem is reduced to solving independent single-channel Schrödinger equations with van der Waals potentials plus centrifugal terms, $W_{nn}(r) = -C_{6}/r^{6} + \hbar^{2}l(l+1)/(2\mu r^{2})$. 
We solve the long-range problem numerically by imposing the standard scattering boundary condition at ($r \to \infty$)
\begin{equation}
\mathbf{\Psi}(r) = \mathbf{F}(r) - \mathbf{G}(r)\mathbf{K}.
\end{equation}
Here, $\mathbf{F}(r)$ and $\mathbf{G}(r)$ are diagonal matrices representing linearly independent solutions. 
For asymptotically open channels ($E \ge E_n$), they behave as $F_n(r) \sim k_n^{-1/2} \hat{j}_l(k_n r)$ and $G_n(r) \sim k_n^{-1/2} \hat{y}_l(k_n r)$ as $r \to \infty$, where $k_n = \sqrt{2\mu(E - E_n)}/\hbar$. 
For asymptotically closed channels ($E < E_n$), the physical wavefunction must vanish. 
We thus set $F_n(r) = 0$ and define $G_n(r)$ as a exponentially decaying solution, $G_n(r) \sim \sqrt{\kappa_n r} K_{l+1/2}(\kappa_n r)$, with $\kappa_n = \sqrt{2\mu(E_n - E)}/\hbar$.

We initialize the diagonal elements of $\mathbf{F}(r)$ and $\mathbf{G}(r)$ at $r_{\mathrm{max}}$ and propagate them inward to $r_0$ using the renormalized Numerov algorithm~\cite{Johnson_1977}. 
The ratio of the numerically propagated solution must match the short-range ratio, $\mathbf{Q}(r_0)\mathbf{\Psi}(r_0) = \mathbf{\Psi}(r_0 - h)$, which leads to the formula for the $K$-matrix
\begin{equation}
\begin{split}
\mathbf{K} = &\left(\mathbf{G}(r_0 - h) - \mathbf{Q}(r_0) \mathbf{G}(r_0)   \right)^{-1} \\
&\times \left( \mathbf{F}(r_0 - h) -\mathbf{Q}(r_0) \mathbf{F}(r_0)   \right).
\end{split}
\end{equation}
The scattering $S$-matrix is then obtained from the open-open block of the $K$-matrix, $\mathbf{S} = (\mathbf{I} - i\mathbf{K}_{oo})^{-1}(\mathbf{I} + i\mathbf{K}_{oo})$. 

\begin{table}
    \centering
    \caption{Molecule-specific computational parameters used in the minimal quantum scattering model.}
    \label{tab:num_params}
    \footnotesize
    \begin{ruledtabular}
    \begin{tabular}{l c c c c c c}
        System & $r_{\mathrm{min}}$ & $r_0$ & $r_{\mathrm{vdW}}$ & $r_{\mathrm{max}}$ & $V_0$ & $\hbar\gamma_0$ \\
         & $(10^6\,a_0)$ & $(a_0)$ & $(a_0)$ & $(10^3\,a_0)$ & $(\mathrm{K})$ & $(\mu\mathrm{K})$ \\
        \colrule
        $^{87}$Rb$^{133}$Cs & $-110$ & 40.0 & 519.0 & 15.0 & $-14.0$ & $14.4$ \\
        $^{23}$Na$^{87}$Rb  & $-11.5$ & 57.1 & 741.4 & 21.4 & $-13.7$ & $19.2$ \\
        $^{23}$Na$^{39}$K   & $-9.76$ & 37.9 & 491.6 & 14.2 & $-55.2$ & $19.2$ \\
        $^{23}$Na$^{40}$K   & $-28.7$ & 38.0 & 493.5 & 14.2 & $-53.9$ & $19.2$ \\
    \end{tabular}
    \end{ruledtabular}
\end{table}

The formally infinite expansion into Fourier components [Eq.~\eqref{eq:wv_exp}] is truncated symmetrically to $n \in [-n_{\mathrm{max}}, n_{\mathrm{max}}]$. 
Because the spacing between Floquet sidebands scales with the modulation frequency $\Omega$, calculations at high frequencies decouple rapidly and require a smaller basis than required for calculations at low frequencies.
We performed systematic convergence tests for extreme limits of the parameter space, covering collision energies from $1\,\text{nK}$ to $1\,\text{K}$ and modulation frequencies from $1\,\text{Hz}$ to $1\,\text{GHz}$.
Progressively increasing the basis size up to $n_{\mathrm{max}}=1000$ confirmed that $n_{\mathrm{max}}=500$ is sufficient; even in the most demanding regime of low energy and low frequency, expanding from $n_{\mathrm{max}}=500$ to $1000$ alters the resulting loss rate by less than $0.1\%$.

The grid limits are typically set to $r_0 \ll r_{\mathrm{vdW}}$ and $r_{\mathrm{max}} \sim 30\,r_{\mathrm{vdW}}$, where $r_{\mathrm{vdW}} = (2\mu C_6/\hbar^2)^{1/4}$ is the characteristic van der Waals length of the collision.
The grid step size $h$ is determined dynamically: we identify the open Floquet channel possessing the highest local kinetic energy at $r_{0}$ to determine the shortest local de Broglie wavelength, and force 15 grid points per such half-wavelength.

For the strength of the complex absorbing potential $\hbar\gamma_0$, following the experiments, we operate in the saturated loss regime.
In this limit, the computed loss rates become completely insensitive to the exact value of the destruction rate $\gamma_0$, naturally recovering the universal loss limit for a continuous-wave trap ($D=1$).
In practice this means $\hbar \gamma_{0} \sim E_{\mathrm{vdW}}$.
The complete set of molecule-specific computational parameters is summarized in Table~\ref{tab:num_params}.

\begin{figure}[b]
    \centering
    \includegraphics[width=\linewidth]{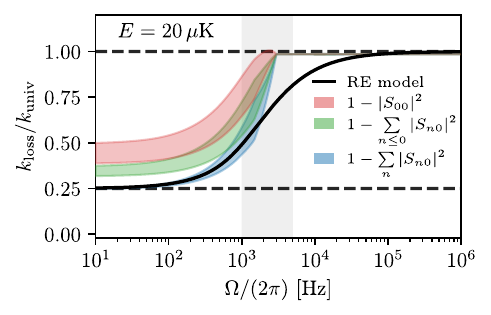}
    \caption{Dependence of the loss rate profile on the definition of $k_{\mathrm{loss}}$ (see text for details). Calculations performed for RbCs at $E=20\mu\mathrm{K}$ for $D=0.25$ (see Fig.~\ref{fig:loss_vs_chop}).}
    \label{fig:loss_def}
\end{figure}

\textit{Appendix B: Loss rate definition.---} In Eq.~\eqref{eq:rate_our_model}, we define loss rate through the defect of the elastic scattering matrix element, $1 - |S_{0,0}|^{2}$.
This definition treats any inelastic transition into a different Floquet sideband ($n\neq 0$) as an effective loss process.
This is justified for shallow ODTs: inelastic transitions that increase the relative kinetic energy of the molecules will cause them to evaporate from the trap.
As shown in Fig.~\ref{fig:loss_def}, relaxing this definition systematically modifies the calculated loss rate at low modulation frequencies.
The alternative definition $1 - \sum\limits_{n \le 0} |S_{n,0}|^2$ (green curve) includes survival in {open} channels that asymptotically lie \textit{above} the incoming threshold ($n \le 0$).
Such transitions do not result in trap heating.
If, on the other hand, survival is defined to include \textit{all asymptotically open} Floquet channels, the resulting loss rate $1 - \sum\limits_{n } |S_{n,0}|^2$ recovers the low-frequency limit of the phenomenological RE model.
Since this sum includes channels that lead to heating of the sample, we choose the definition in Eq.~\eqref{eq:rate_our_model} as the most realistic description of the measured loss rates.

\color{black}

\bibliography{apssamp}% Produces the bibliography via BibTeX.

\end{document}